\documentclass{article}
\usepackage{frascatiphys}
\usepackage{graphicx}
\usepackage{amsmath}

\def\EmissT{\,{}/ \hspace{-1.5ex}  E_{T}}
\begin{document}
\title{ 
FLAVOUR PHYSICS FROM PRESENT TO FUTURE COLLIDERS
}
\author{
Monika Blanke       \\
{\em Institute for Nuclear Physics (IKP) and Institute for Theoretical Particle Physics (TTP),}
\\
{\em Karlsruhe Institute of Technology (KIT), D-76128 Karlsruhe, Germany} 
}
\maketitle
\baselineskip=10pt
\begin{abstract}
In these proceedings we provide a brief overview of the status of flavour physics, with focus on opportunities to discover New Physics in flavour-violating decays at current and future colliders.
\end{abstract}
\baselineskip=14pt

\section{Introduction}

At the end of run 2 of the Large Hadron Collider (LHC), still no evidence for the production of new particles has been found. The resulting limits on the mass scale of New Physics (NP) in many cases reach beyond $1\,\text{TeV}$, and we have to face the  possibility that the energy reached at the LHC might not be sufficient for a direct NP discovery. Although it is certainly too early to draw such depressing conclusions, the alternative indirect probes of NP in low-energy precision observables become increasingly relevant. While the reach of precision tests of the electroweak sector is limited to about $10\,\text{TeV}$, flavour-changing neutral current (FCNC) processes are sensitive to much higher scales, $1000\,\text{TeV}$ and beyond\cite{Buras:2014zga}. In these proceedings we recapitulate the opportunities to discover NP in flavour observables, paying particular attention to some tensions in the data that arose over the past years.

\section{New Physics Opportunities with the Unitarity Triangle}

Consistency checks of the CKM mechanism in terms of Unitarity Triangle (UT) fits have a long tradition\cite{Charles:2004jd} and tell a story of success of the Standard Model (SM). A drawback of these global analyses is, however, that emerging tensions in particular channels remain hidden due to the large number of observables entering the fit. A good alternative hence is to compare the data on a few specific FCNC observables with their predictions using the CKM matrix elements determined from tree-level charged-current decays.

\begin{figure}
\centering{\includegraphics[width=.55\textwidth]{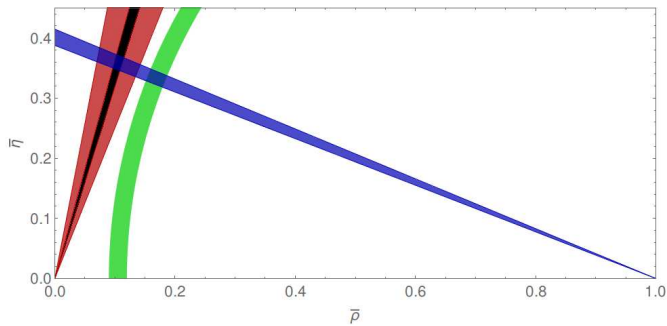}}
\caption{\label{fig1}Constraints on the Unitarity Triangle from the measurement of $\sin2\beta$ (blue), the ratio $\Delta M_d/\Delta M_s$ (green), and the tree-level determination of the angle $\gamma$ (red). The future expected $1^\circ$ sensitivity for $\gamma$ by LHCb and Belle II is shown in black.
Figure taken from Ref.\cite{Blanke:2018cya}.
}
\end{figure}

Unfortunately a precise determination of the full reference unitarity triangle\cite{Grossman:1997dd} is still impeded due to the persisting $|V_{ub}|$ problem\cite{Bouchard:2019all}; yet an interesting tension arises already in the current data, shown in Figure \ref{fig1}. The red area displays the LHCb measurement\cite{Kenzie:2319289} of the UT angle $\gamma$ in tree-level $B\to DK$ decays, and the expected future $1^\circ$ precision\cite{Cerri:2018ypt} by both LHCb and Belle II is indicated by the black line. One can see that such large values for $\gamma>70^\circ$ are inconsistent with its indirect determination through the ratio of mass differences $\Delta M_d/\Delta M_s$ in $B_{d,s}-\bar B_{d,s}$ mixings\cite{Blanke:2018cya,Blanke:2016bhf}, shown by the green band, irresepective of the size of $|V_{ub}|$. Such tension, if confirmed by future more accurate data would be an unambiguous sign of NP contributions to $\Delta M_d$ and/or $\Delta M_s$.

In addition to the reduced uncertainty in the measurement of $\gamma$, a crucial ingredient to unravelling this potential anomaly is the impressive improvement made in the theoretical determination of the $SU(3)_F$-violating ratio $\xi$ of the hadronic matrix elements entering $B_{d,s}-\bar B_{d,s}$ mixing both by lattice QCD\cite{Bazavov:2016nty} and QCD sum rule\cite{Grozin:2016uqy} calculations. The ball is now in the field of LHCb and Belle II to improve the tree-level determination of $\gamma$ and thereby confirm whether indeed the ratio $\Delta M_d/\Delta M_s$ is affected by NP contributions.

\section{Lepton Flavour Universality Violating Anomalies}

Over the past years, several anomalies in both charged and neutral current semileptonic $B$ meson decays emerged in the data. Interestingly, both of these sets of ``$B$ anomalies'' are related to the violation of lepton flavour universality (LFU).

\subsection{The Charged Current $b\to c\tau\nu$ Transitions}

The ratios
\begin{equation}
R(D^{(*)})=\frac{\text{BR}(B\to D^{(*)} \tau
  \nu)}{\text{BR}(B\to D^{(*)} \ell \nu)} \qquad (\ell=e,\mu)
\end{equation}
provide a clean test of LFU in charged current $b\to c$ transitions, mediated in the SM by tree-level $W^\pm$ boson exchange.
Various measurements by BaBar\cite{Lees:2012xj}, Belle\cite{Huschle:2015rga} and LHCb\cite{Aaij:2015yra} indicate an enhancement with respect to the SM prediction, with the current HFLAV combination\cite{Amhis:2016xyh} finding a $3.1\sigma$ anomaly. An experimental consistency check of this result will be provided by a measurement of the corresponding baryonic ratio
\begin{equation}
R(\Lambda_c)=\frac{\text{BR}(\Lambda_b \to \Lambda_c \tau
  \nu)}{\text{BR}(\Lambda_b \to \Lambda_c \ell \nu)} \qquad (\ell=e,\mu)
\end{equation}
which is predicted model-independently to be\cite{Blanke:2018yud}
\begin{equation}
R(\Lambda_c) \simeq R(\Lambda_c)_\text{SM} \left( 0.262 \frac{{R}(D)}{{R}_{\rm SM}(D)} + 0.738 \frac{{R}(D^*)}{{R}_{\rm
      SM}(D^*)} \right)
=0.38\pm0.01\pm0.01\,.
\end{equation}

Potential NP contributions at the origin of this anomaly can be systematically described by the effective Hamiltonian
\begin{equation}
 {\cal H}_{\rm eff}(b\to c\tau\nu)=  2\sqrt{2} G_{F} V^{}_{cb} \big[(1+C_{V}^{L}) O_{V}^L +   C_{S}^{R} O_{S}^{R} 
   +C_{S}^{L} O_{S}^L+   C_{T} O_{T}\big] \,.
\label{Heff}
\end{equation}
Several groups\cite{Blanke:2018yud,Murgui:2019czp,Aebischer:2019mlg} have fitted the Wilson coefficients $C_i$ to the available data. 

Matching the effective Hamiltonian to simplified NP models in which the $b\to c\tau\nu$ transition arises from the tree-level exchange of a single mediator, a number of different scenarios emerges. Relevant contributions from a heavy charged $W'$ gauge boson\cite{He:2012zp} are challenged both by electroweak precision constraints\cite{Feruglio:2017rjo} and by high-$p_T$ di-$\tau$ data at the LHC\cite{Faroughy:2016osc}. Charged Higgs contributions\cite{Kalinowski:1990ba}
generate a large branching ratio $\text{BR}(B_c\to\tau\nu)>50\%$ and are put under pressure by mono-$\tau$ searches\cite{Greljo:2018tzh}. The best option for a NP explanation of the anomaly hence remains a scalar or vector leptoquark, see e.\,g.\ Refs.\cite{Freytsis:2015qca,Alonso:2015sja,Bordone:2017bld,Calibbi:2015kma}.

Further insight on the underlying NP can be obtained by measuring differential and angular observables\cite{Blanke:2018yud,Nierste:2008qe} which can discriminate between the different scenarios. Additionally, decay modes related to $b\to c\tau\nu$ by $SU(2)_L$ symmetry, like $B\to K^{(*)}\nu\bar\nu$, $B_s\to\tau^+\tau^-$, $B\to K^{(*)}\tau^+\tau^-$, $\Upsilon\to\tau^+\tau^-$ and $\psi\to\tau^+\tau^-$ can receive significant NP contributions\cite{Calibbi:2015kma,Crivellin:2017zlb,Aloni:2017eny}, depending on the NP model at work, and already now challenge some of the existing models. Overall, due to the large number of complementary observables, a NP origin of the anomaly can unambiguously be tested in both high-$p_T$ and low-energy flavour data.

\subsection{The Neutral Current $b\to s\ell\ell$ Modes}

An equally interesting set of anomalies has appeared in measurements of $B$ decays mediated by $b\to s\ell\ell$. The most relevant deviations from the SM are seen in the angular distribution of $B \to K^*\mu^+\mu^-$\cite{Aaij:2015oid}, as well as in the LFU ratios\cite{Aaij:2017vbb}
\begin{equation}
R_{K^{(*)}} = \frac{\text{BR}(B\to K^{(*)}\mu^+\mu^-)}{\text{BR}(B\to K^{(*)}e^+e^-)}\,.
\end{equation}
Again, potential NP effects can conveniently be described as contributions to the Wilson coefficients in
\begin{equation}
\mathcal{H}_\text{eff}(b\to s\ell\ell)= -\frac{4 G_F}{\sqrt{2}} V_{tb}^* V_{ts} \frac{e^2}{16\pi^2}\sum_i(C_i {\cal O}_i +C'_i {\cal O}'_i)+h.c.\,.
\end{equation}
Here, the terms most sensitive to NP are the magnetic dipole operators ${\cal O}^{(\prime)}_7$ and the four-fermion operators ${\cal O}^{(\prime)}_{9,10}$. Note that the latter can be generated at tree level by $Z'$\cite{Altmannshofer:2013foa} or leptoquark\cite{Alonso:2015sja,Calibbi:2015kma,Hiller:2014ula} exchanges but are loop-suppressed in the SM, turning them into sensitive probes of NP.

Currently, one of the most promising solutions to the anomaly is a NP scenario with purely left-handed couplings, generating\cite{Aebischer:2019mlg} (see also Ref.\cite{Alguero:2019ptt} for recent global fits) 
\begin{equation}\label{C9-C10}
\delta C_{9}^{bs\mu\mu}=-\delta C_{10}^{bs\mu\mu} \simeq -0.53\,.
\end{equation}
This scenario can accomodate a suppression of ${\rm BR}(B_s\to\mu^+\mu^-)$ with respect to its SM value, and is easy to realise in concrete NP scenarios. 

Among the most popular NP models for this anomaly is a TeV-scale $SU(2)_L$ singlet vector leptoquark coupling dominantly to left-handed quarks and leptons. Not only can it generate the required NP contribution in \eqref{C9-C10} without generating unwelcome effects in $B_s-\bar B_s$ mixing, but it can simultaneously also accomodate the required NP effect in the charged current transition $b\to c\tau\nu$.
Interestingly such a particle arises from the Pati-Salam gauge symmetry unifying quarks and leptons\cite{Pati:1974yy}. Following this observation various model-building attempts\cite{DiLuzio:2017vat,Blanke:2018sro} have been undertaken to construct a viable UV-complete model for the $B$ decay anomalies.

Instead of dwelling further on the model-building challenges, we turn our attention to complementary probes of such a NP explanation of the anomalies. In $B$ physics, important tests are given by LFU violating angular observables in $B\to K^*\ell^+\ell^-$\cite{Capdevila:2016ivx}, the $SU(2)_L$-related modes $B\to K^{(*)}\nu\bar\nu$, $B_s\to\tau^+\tau^-$, $B\to K^{(*)}\tau^+\tau^-$\cite{Calibbi:2015kma,Crivellin:2017zlb}, and the lepton flavour violating meson decays $B\to K^{(*)}\tau^\pm\mu^\mp$ and $B_s\to \tau^\pm\mu^\mp$\cite{Bordone:2018nbg}. In the lepton sector, these NP scenarios and their flavour structure are probed by lepton flavour violating $\mu$ and $\tau$ decays\cite{Crivellin:2017zlb,Blanke:2018sro,Bordone:2018nbg,Barbieri:2019zdz}. Last but not least, also the high-energy frontier places important constraints on these scenarios, both in terms of the direct production of the vector leptoquark and its partner states\cite{Baker:2019sli} and in high-$p_T$ di-lepton tails\cite{Greljo:2017vvb}.

\section{\boldmath High-$p_T$ Routes to Flavour}

In addition to high-precision measurements of flavour-violating meson decays, the NP flavour structure can also be explored at the high-energy frontier, with ample opportunities at the High Luminosity phase of the LHC (HL-LHC) and future lepton or hadron colliders.

With the current hints for anomalies in flavour-violating $B$ decays, it is conceivable that the underlying NP, coupling dominantly to the third generation, also leaves an observable imprint on flavour-violating top quark couplings.  While the current bounds on transitions like $t\to (c,u)h$, $t\to (c,u)\gamma$ and $t\to (c,u)Z$ are too weak to put relevant limits on concrete NP models, the situation will significantly improve at the HL-LHC and in particular at a future high-energy hadron collider, due to the large number of top quarks produced\cite{Mandrik:2018yhe}.

\begin{figure}
\centering{\includegraphics[width=.6\textwidth]{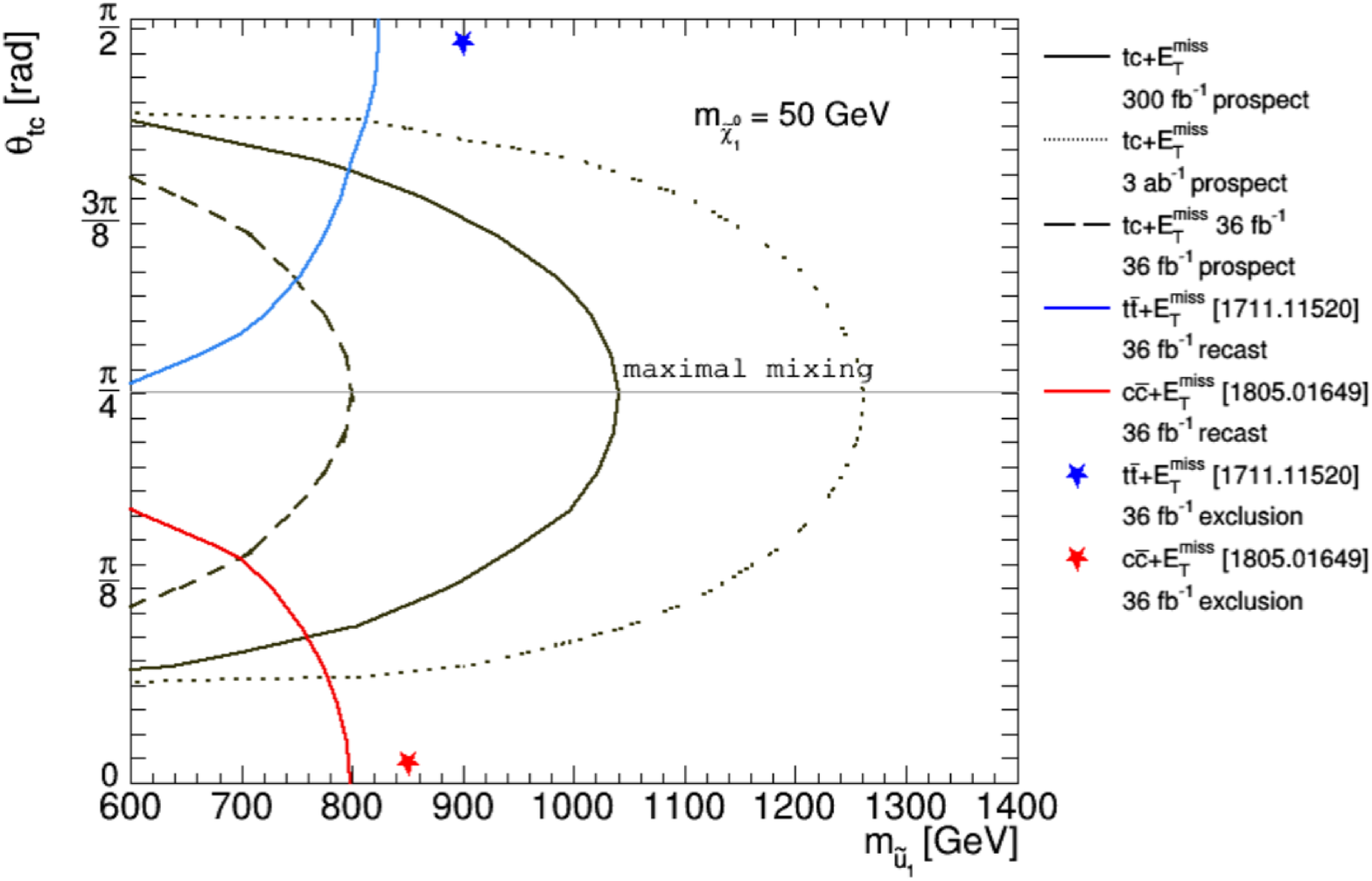}}
\caption{\label{fig2}Bounds on the mass of the lightest squark, assuming it to be a mixture of stop and scharm flavour eigentstates. The blue and red curves display the constraints from  $t\bar t+\EmissT$ and $c\bar c+\EmissT$, respectively, as a function of the mixing angle $\theta_{tc}$. The black curves indicate the expected reach of a dedicated search for $tc+\EmissT$. Figure taken from Ref.\cite{Chakraborty:2018rpn}.
}
\end{figure}

If the scale of NP is low enough that the new particles can directly be produced, then their flavour structure has an immediate impact on their decay products. For instance, in the case of supersymmetric (SUSY) models, the  presence of flavour mixing affects the accessible final states for squark pair production and therefore alters the corresponding phenomenology: The presence of mixing between the top and the charm squark significantly weakens the constraints from squark searches assuming flavour conservation\cite{Blanke:2013zxo,Chakraborty:2018rpn}. At the same time, the flavour-violating final state $tc+\EmissT$ becomes relevant\cite{Blanke:2013zxo,Chakraborty:2018rpn,Han:2003qe}, for which a dedicated search would be a promising way to discover scenarios with a large stop-scharm mixing angle, see Figure \ref{fig2}. Note that the $tc+\EmissT$ signature can arise also in other NP scenarios, like top-flavoured dark matter with a non-minimal flavour structure\cite{Blanke:2017tnb}. Interestingly in the latter case the cross-section can be large even if the relevant flavour mixing angles are zero.

\section{Summary and Outlook}

In these proceedings we provided a brief overview of the opportunities to discover NP in flavour-violating observables at present and future collider experiments. We did not cover charm decays here which, while experimentally a very interesting and rich field, still constitute a major problem for precise theoretical predictions due to the dominance of long-distance effects. We did not discuss kaon decays either, despite their unique sensitivity to NP contributions from very high energy scales, as the exploration of this exciting field does mostly not involve collider experiments. Recent reviews of the status of kaon physics, including the discussion of a potential anomaly in $\epsilon'/\epsilon$, can be found in Ref.\cite{Buras:2018wmb}.

\section{Acknowledgements}
I thank the organisers of LFC19 for the kind invitation and the STRONG-2020 network for the financial support that allowed me to contribute to this stimulating workshop. This work was partially supported by the Deutsche Forschungsgemeinschaft (DFG, German Research Foundation) under grant  396021762 - TRR~257.


%

%
\end{document}